\documentclass[english,12pt,a4]{article}
\usepackage[T1]{fontenc}
\usepackage[latin1]{inputenc}
\usepackage{a4}
\usepackage{amsmath}
\usepackage{graphicx}

\makeatletter

\providecommand{\LyX}{L\kern-.1667em\lower.25em\hbox{Y}\kern-.125emX\@}

 \newcommand{\lyxaddress}[1]{
   \par {\raggedright #1 
   \vspace{1.4em}
   \noindent\par}
 }

\makeatother
\begin{document}

\title{\textbf{\Large Antiperiodic solitons of the Goldstone model on $S^{1}$}}

\author{{\normalsize C G Doudoulakis}}

\date{{}}

\maketitle

\lyxaddress{Department of Physics and Institute of Plasma Physics, University
of Crete, Heraklion, Greece}

\begin{abstract}
Our purpose is to present all static solutions of the Goldstone model
on a circle in 1+1 dimensions with an antiperiodicity condition imposed
on the scalar fields. Jacobi elliptic and standard trigonometric functions
are used to express the solutions found and stability analysis of
the latter is what follows. Classically stable quasi-topological solitons
are identified.
\end{abstract}

\section{{\normalsize Introduction}}

The purpose of this note is to present the complete list of static
classical solutions of the Goldstone model on a circle $S^{1}$ of
radius $L$ and the corresponding bifurcation tree together with a
study of the stability of our solutions. It comes as a supplement
to a previous note {[}1{]} which was searching for such solutions
but with periodicity condition imposed there. Many of the results
have obvious similarities with those in {[}1{]}. We notice those and
make comparisons with this reference. We also check the case of mixed
boundary conditions and place a note with our conclusions in the end.
The following analysis has both mathematical {[}8{]} and physical
interest as it can be useful for the search of stable solitons in
the two-Higgs standard model (2HSM) or the minimal supersymmetric
standard model.

\section{{\normalsize The classical solutions}}

The Langrangian of our model is\begin{equation}
\mathcal{L}=\frac{1}{2}(\partial _{\mu }\phi _{1})^{2}+\frac{1}{2}(\partial _{\mu }\phi _{2})^{2}-V(\phi _{1},\phi _{2})\: ,\quad \mu =0,1\end{equation}
 with $\phi _{1}$ and $\phi _{2}$ real Higgs fields and potential
of the following form:

\begin{equation}
V(\phi _{1},\phi _{2})=\frac{1}{4}(\phi _{1}^{2}+\phi _{2}^{2}-1)^{2}\end{equation}
 We impose antiperiodic boundary conditions on the scalar fields with
coordinates $x\in [0,2\pi L].$

\begin{equation}
\phi _{i}\left(x+2\pi L\right)=-\phi _{i}\left(x\right)\: ,\quad i=1,2\end{equation}
The general static solution of model (1) can be expressed in terms
of Jacobi elliptic functions:

\begin{equation}
\phi _{1}=\sqrt{R}\: \cos \Omega ,\quad \phi _{2}=\sqrt{R}\: \sin \Omega \end{equation}

\begin{equation}
R(x)=a_{1}+a_{2}sn^{2}(\sqrt{2}\: \Lambda (x-x_{0})\: ,k)\end{equation}

\begin{equation}
\Omega (x)=C\! \int _{\xi }^{x}\! \frac{1}{R(y)}dy\end{equation}
where $\! C\left(a_{1}\left(k,\Lambda \right),a_{2}\left(k,\Lambda \right)\right),\! x_{0},\! \xi $
are constants while $sn$ denotes the Jacobi elliptic function; $sn(z,k),$
$sn^{2}(z,k)$ are periodic functions on the real axis with periods
$4K(k)$ and $2K(k)$, respectively. $K(k)$ is the complete elliptic
integral of the first kind. Inserting (4)-(6) into the equations corresponding
to (1) leads to several conditions on the parameters and as a consequence
to three different types of non-trivial solutions. We start with the
simplest, the trivial solution.

\subsection{{\normalsize The trivial solution}}

The energy functional for static configurations is given by \begin{equation}
E=\int _{0}^{2\pi L}dx\: [\frac{1}{2}(\frac{\partial \phi _{1}}{\partial x})^{2}+\frac{1}{2}(\frac{\partial \phi _{2}}{\partial x})^{2}+V(\phi _{1},\phi _{2})].\end{equation}
Apart from the vacuum solutions which have $E_{vac}=0$, one can immediately
think of the simplest solution, the trivial one

\begin{equation}
\phi _{1}=\phi _{2}=0\; ,\; E_{0}=\frac{L\pi }{2}\end{equation}
which exists for all values of $L$. The corresponding small oscillation
eigenmodes, labeled by $j$, have

\begin{equation}
\tilde{\omega }^{2}(j)=\frac{1}{L^{2}}(\, \left(j+1/2\right)^{2}-L^{2})\: ,\quad j=0,1,2...\end{equation}
 This solution is stable until $L=1/2$ because $\tilde{\omega }^{2}(0)<0$
for $L>1/2$. Many additional solutions, some of which were discussed
in {[}2,3{]} bifurcate from the solution $\phi _{1}=\phi _{2}=0$
at critical values of $L$.

\subsection{{\normalsize Three types of non-trivial solutions}}

Now, we present without many details the three non-trivial solutions
and mention their similarities with those in {[}1{]}. The simplest
case one can think, is to set $a_{2}=0$ so that from (5) we obtain
$R(x)=a_{1}$ and the solution becomes

\begin{equation}
\phi _{1}=\sqrt{1-\frac{\left(N+\frac{1}{2}\right)^{2}}{L^{2}}}\! \cos \left(\frac{\left(N+\frac{1}{2}\right)x}{L}\right)\end{equation}

\begin{equation}
\phi _{2}=\sqrt{1-\frac{\left(N+\frac{1}{2}\right)^{2}}{L^{2}}}\! \sin \left(\frac{\left(N+\frac{1}{2}\right)x}{L}\right)\end{equation}
where $N$ is an integer. This solution is called type-I. One can
observe that the above solution reduces to (8) when $L=N+1/2$ i.e.
when one of the $\tilde{\omega }$ of equation (9) crosses zero. The
Higgs field winds $(2N+1)/2$ times around the top of the Mexican
hat. It's energy is given by

\begin{equation}
E_{I}\left(L,N\right)=\frac{\pi }{L}\left(N+\frac{1}{2}\right)^{2}-\frac{\pi }{2L^{3}}\left(N+\frac{1}{2}\right)^{4}.\end{equation}
 If we denote the above result which corresponds to the antisymmetric
case we study here with $E_{I}^{A}$ and the result of {[}1{]} for
the same type of solutions with $E_{I}^{S}$ where $S$ means {}``Symmetric''
and $A$ {}``Antisymmetric'' then we notice that \begin{eqnarray*}
E_{I}^{A}\left(N\, ,L\right) & = & E_{I}^{S}\left(N+\frac{1}{2}\: ,L\right)
\end{eqnarray*}

Another choice is to set $a_{1}=0$ on (5) so the solution now involves
the Jacobi elliptic functions. $C$ becomes zero as well and the solution
is

\begin{equation}
\phi _{1}=2k\Lambda sn\left(\sqrt{2}\Lambda x,\! k\right)\: ,\quad \quad \phi _{2}=0\: ,\quad \quad \Lambda ^{2}=\frac{1}{2\left(1+k^{2}\right)}\end{equation}
 This solution is called type-II. In fact, it corresponds to an oscillation
of the Higgs field in the $\phi _{2}=0$ plane about the origin $\phi _{1}=0=\phi _{2}$.
If we take account of the antiperiodicity condition (3), then the
argument $k$ of the Jacobi elliptic function is related to the radius
$L$ of $S^{1}$ through the following formula

\includegraphics[  scale=0.55]{picture1}

\begin{equation}
L=\frac{\left(2m+1\right)K(k)}{\pi }\sqrt{1+k^{2}}\end{equation}
 for some integer $m$. When we reach the limit of $k\rightarrow 0$
(i.e. $L\rightarrow m+\frac{1}{2}$) the solution (13) approaches
(8). The energy of the solution above is given by the integral 

\begin{equation}
E_{II}=\frac{4\left(2m+1\right)}{\sqrt{2}\Lambda \left(1+k^{2}\right)^{2}}\! \! \int _{0}^{K(k)}dy\! \left[\left(k^{2}sn^{2}\left(y,k\right)-\frac{1+k^{2}}{2}\right)^{2}+\frac{2k^{2}-1-k^{4}}{8}\right]\end{equation}
 and by means of Elliptic integrals it becomes

\[
E_{II}=\frac{4\left(2m+1\right)}{\sqrt{2}\Lambda \left(1+k^{2}\right)^{2}}\; \left[\frac{K(k)}{24}\: \left(3k^{4}+2k^{2}-5\right)+\frac{E(k)}{3}\: \left(k^{2}+1\right)\right]\]
Two specific values of $k$ with the corresponding results follow

for $k=0$

\begin{equation}
E_{II}\left(L=m+1/2,m\right)=\frac{\left(2m+1\right)\pi }{4}\end{equation}

for $k=1$

\begin{equation}
E_{II}\left(L=\infty ,m\right)=\frac{4\left(2m+1\right)}{3\sqrt{2}}\end{equation}
 The comparison of $E_{II}^{A}$ which represents our solutions, with
$E_{II}^{S}$ which represents the symmetric case of the same type
of solutions studied in {[}1{]}, implies that \begin{eqnarray*}
E_{II}^{A}(m) & = & E_{II}^{S}(m+\frac{1}{2})
\end{eqnarray*}

If none of $a_{1},a_{2}$ are zero then we are led to type-III solutions
where we have the following conditions for $a_{1},a_{2}$ and $C^{2}$

\begin{equation}
a_{1}=\frac{2}{3}\left(1-2\Lambda ^{2}\left(1+k^{2}\right)\right),\qquad a_{2}=4k^{2}\Lambda ^{2}\end{equation}

\begin{align}
C^{2}= & \frac{4}{27}\left(1+\left(4k^{2}-2\right)\Lambda ^{2}\right)\left(1+\left(4-2k^{2}\right)\Lambda ^{2}\right)\left(1-2\Lambda ^{2}\left(1+k^{2}\right)\right)\nonumber \\
= & \frac{2}{9}\left(1+\left(4k^{2}-2\right)\Lambda ^{2}\right)\left(1+\left(4-2k^{2}\right)\Lambda ^{2}\right)a_{1}
\end{align}

Here, one can explicitly observe the fact that when $a_{1}=0$ then
$C^{2}=0$ as well. In addition, $R$ and $C^{2}$ must not be negative.
This means another condition 

\begin{equation}
\Lambda ^{2}\leq \frac{1}{2\left(1+k^{2}\right)}\end{equation}
 with the equality leading to type-II solutions.

In order for $\Omega $ to satisfy $\Omega \left(x+2\pi L\right)=\Omega \left(x\right)+\pi $
and $R$ to be periodic (so for the solution to be antiperiodic as
we want) on $[0,2\pi L]$ the following equations must hold:

\begin{equation}
C\! \int _{0}^{2\pi L}\frac{1}{R(y)}dy=\left(2n+1\right)\pi \end{equation}

\begin{equation}
L=\frac{\left(2m+1\right)K(k)}{\sqrt{2}\pi \Lambda }\end{equation}
 where $m,n$ are positive integers which respectively determine the
number of oscillations of the modulus of the Higgs field and the number
of the Higgs field windings around the origin $\phi _{1}=0=\phi _{2}$
in a period $2\pi L$.

Now we return to (18) and (19) and analyze further type-III solutions.
Solving these equations for $k=0$, remembering that $K(0)=\pi /2$,
we find the critical values of $L$ where these solutions start to
exist:

\begin{equation}
\Lambda ^{2}=\frac{m^{2}}{6\left(2n+1\right)^{2}-4m^{2}}\: \: \Rightarrow \: \: L^{2}=\frac{3}{4}\left(2n+1\right)^{2}-\frac{m^{2}}{2}\end{equation}
 The expression for $\Lambda ^{2}$ together with (20) leads to the
condition $2n+1>m$ on the integers $m$ and $n$ ($m$ and $n$ $\neq 0$)
. For any $n$ and $m\neq 2n+1$ the coefficient of $sn^{2}$ in the
integral

\includegraphics[  scale=0.55]{picture2}

(6) is proportional to $k^{2}$ and if one expands the solution in
powers of $k^{2}$ is led to the following formulae

\begin{equation}
\Lambda ^{2}=\frac{m^{2}}{6\left(2n+1\right)^{2}-4m^{2}}\! \! \left(1+\frac{k^{2}}{2}\right)+\mathcal{O}\left(k^{4}\right)\end{equation}

\begin{equation}
L^{2}=\frac{3}{4}\left(2n+1\right)^{2}-\frac{m^{2}}{2}+\mathcal{O}\left(k^{4}\right)\end{equation}

\begin{equation}
C^{2}=\frac{4}{3}\frac{\left(\left(2n+1\right)^{2}-m^{2}\right)^{2}\left(2n+1\right)^{2}}{\left(3\left(2n+1\right)^{2}-2m^{2}\right)^{3}}+\mathcal{O}\left(k^{4}\right)\! .\end{equation}
This expansion is necessary as there is no closed form for integral
(21). In the limit $k=0$ we observe that the $2n$ solutions of type-III
yield the type-I solution with $N=n+1/2$ at $L^{2}=\left(3/4\right)\left(2n+1\right)^{2}-m^{2}/2$
, where $m=1,2,...,2n$. For fixed values of $k,\, n,\, m$ we find
a single solution $\Lambda ^{2}\left(k,\: \left(2n+1\right)/m\right)$
obeying the following properties

\begin{equation}
\Lambda ^{2}\left(k=0,\: \left(2n+1\right)/m\right)=\frac{m^{2}}{6\left(2n+1\right)^{2}-4m^{2}}\: ,\quad \quad \Lambda ^{2}\left(k=1,\: \left(2n+1\right)/m\right)=1/4\end{equation}
This is illustrated in figure 1 for two different values of the pair
$(n,m)$ by the solid curves, the dashed curve representing the limit
(20). 

The energy of the general type-III solution is given by the integral

\begin{align}
E_{III}= & \frac{2m+1}{\sqrt{2}\: \Lambda }\, \nonumber \\
\cdot  & \int _{0}^{K(k)}dy\: \left[\left(a_{1}-1+a_{2}sn^{2}(y,k)\right)^{2}+\frac{1}{6}\: \left(1+16\Lambda ^{4}\left(k^{2}-1-k^{4}\right)\right)\right]
\end{align}
which, by means of $K(k),\: E(k)$ becomes

\begin{align*}
E_{III}= & \frac{2m+1}{\sqrt{2}\: \Lambda }\: \left[K(k)\: \left(\frac{8}{9}\, \Lambda ^{2}k^{2}\left(1+\Lambda ^{2}-\Lambda ^{2}k^{2}\right)-\frac{8}{9}\, \Lambda ^{2}\left(2+\Lambda ^{2}\right)+\frac{5}{18}\right)+\frac{8}{3}\, \Lambda ^{2}E(k)\right]
\end{align*}
For $k=1$ one obtains

\begin{equation}
E_{III}\left(1,m,n\right)=\frac{4\left(2m+1\right)}{3\sqrt{2}}=E_{II}\left(1,m\right).\end{equation}
 This shows that solution III approaches solution II in the limit
$k\rightarrow 1$, a fact which also appears in {[}1{]} .

The energies of the solutions of some low-lying branches are presented
in figure 2 as functions of $L$ . For $L>5/2$ all four types of
solutions coexist and satisfy

\begin{equation}
E_{I}\left(L,N=0\right)<E_{I}\left(L,N=1\right)<E_{III}\left(L,m=1,n=1\right)<E_{II}\left(L,m=1\right)<E_{0}\left(L\right).\end{equation}
 The four stars on the upper part of the figure show the position
of the four bifurcation points $L=3.279,$ $3.775,4.093,4.272$ of
the $n=2,\: m=3,2,1,0$ solutions respectively from the $N=2$ type-I
solution. The two lower stars show the bifurcation values $\left(L=2.179,\: 2.5\right)$
of the $n=1,\: m=2,1$ solutions respectively from the $N=1$ type-I
solution.

\section{{\normalsize Stability Analysis}}

\subsection{{\normalsize Type-I solutions}}

In order to analyze the stability of type-I solutions we follow the
steps done in {[}1{]} which uses notions that can be found in {[}10{]}
and {[}15{]}. Thus perturbing the fields $\phi _{a}\: ,a=1,2$ around
the classical solution (10)-(11), denoted here by $\phi _{a}^{cl}$
,

\begin{equation}
\phi _{a}\left(x\right)=\phi _{a}^{cl}\left(x\right)+\eta _{a}\left(x\right)\exp \left(-i\omega t\right)\end{equation}
 we are led to the following equation for the normal modes:

\begin{equation}
A\left(\tilde{N},L\right)\left(\begin{array}{c}
 \eta _{1}\\
 \eta _{2}\end{array}
\right)=\omega ^{2}\left(\begin{array}{c}
 \eta _{1}\\
 \eta _{2}\end{array}
\right)\end{equation}

\begin{equation}
A\left(\tilde{N},L\right)\equiv -\frac{d^{2}}{dx^{2}}+2\left(1-\left(\frac{\tilde{N}}{L}\right)^{2}\right)\left(\begin{array}{cc}
 c^{2} & sc\\
 sc & s^{2}\end{array}
\right)-\left(\frac{\tilde{N}}{L}\right)^{2}\end{equation}
 where $\omega ^{2}$ is the eigenvalue and $c=\cos \left(\tilde{N}x/L\right)$
, $s=\sin \left(\tilde{N}x/L\right)$ , $N+1/2\equiv \tilde{N}$.

The complete list of eigenvalues of the operator $A\left(\tilde{N},L\right)$
for $\tilde{N}\geq 1$ can be obtained by classifying its invariant
subspaces. This can be done by using the Fourier decomposition. Type-I
solution has a twisted translational invariance and one can check
that for an integer $n\geq \tilde{N}$ the following finite-dimensional
vector spaces%
\footnote{These vector spaces should have been the same as in (34) of {[}1{]}
(with $N\rightarrow \tilde{N}$) but they are not, due to a misprint
in {[}1{]}. Here we correct this by writing these spaces explicitly
in equations (34) and (35).%
} are preserved by $A\left(\tilde{N},L\right)$ 

\begin{equation}
V_{n}=Span\left\{ a_{p}\cos \frac{px}{L}\: ,\: \beta _{p}\sin \frac{px}{L}\: ,\: p-n=0\left(mod2\tilde{N}\right)\: ,\: \left|p\right|\leq n\right\} ,\end{equation}

\begin{equation}
\tilde{V_{n}}=Span\left\{ a_{p}\cos \frac{px}{L}\: ,\: -\beta _{p}\sin \frac{px}{L}\: ,\: p-n=0\left(mod2\tilde{N}\right)\: ,\: \left|p\right|\leq n\right\} \end{equation}
 under the following condition

\begin{equation}
a_{p}=\beta _{p}\qquad \quad \quad if\qquad p-n+2\tilde{N}>0\end{equation}
 where $a_{p}$ , $\beta _{p}$ are arbitrary constants. The operator
$A\left(\tilde{N},L\right)$ can then be diagonalized on each of the
finite-dimensional vector spaces above, leading to a set of algebraic
equations.

To be more specific, define $\lambda _{1}\equiv 2\left(1-\lambda _{2}\right)$,
$\lambda _{2}\equiv \left(\tilde{N}/L\right)^{2}$. Also, consider
the vector

\[
\left(\begin{array}{c}
 V_{\tilde{N}+k}\\
 V_{\tilde{N}-k}\end{array}
\right)\quad \quad \qquad with\quad \; \: V_{\tilde{N}+k}\equiv \left(\begin{array}{c}
 \cos \left(\frac{\tilde{N}+k}{L}\right)x\\
 \sin \left(\frac{\tilde{N}+k}{L}\right)x\end{array}
\right)\: ,\: V_{\tilde{N}-k}\equiv \left(\begin{array}{c}
 \cos \left(\frac{\tilde{N}-k}{L}\right)x\\
 \sin \left(\frac{\tilde{N}-k}{L}\right)x\end{array}
\right)\]
 Acting with the operator $A\left(\tilde{N},L\right)$ on one of the
above vectors (say $V_{\tilde{N}+k}$) we have the following steps:

\[
\left\{ -\frac{d^{2}}{dx^{2}}+\frac{\lambda _{1}}{2}\: \left(\begin{array}{cc}
 2c^{2} & 2sc\\
 2sc & 2s^{2}\end{array}
\right)-\lambda _{2}-\omega ^{2}\right\} V_{\tilde{N}+k}=0\quad \Rightarrow \]

\includegraphics[  scale=0.55]{picture3}

\[
\Rightarrow \left\{ -\frac{d^{2}}{dx^{2}}+\frac{\lambda _{1}}{2}\, \mathbf{1}-\lambda _{2}-\omega ^{2}+\frac{\lambda _{1}}{2}\, \mathbf{M}\right\} V_{\tilde{N}+k}=0\, .\]
 Where $\mathbf{1}$ is the $2\times 2$ unit matrix and 

\[
\mathbf{M}\equiv \left(\begin{array}{cc}
 \cos \left(\frac{2\tilde{N}x}{L}\right) & \sin \left(\frac{2\tilde{N}x}{L}\right)\\
 \sin \left(\frac{2\tilde{N}x}{L}\right) & -\cos \left(\frac{2\tilde{N}x}{L}\right)\end{array}
\right)\]
 which has the action

\[
\mathbf{M}V_{\tilde{N}+k}=V_{\tilde{N}-k}\: .\]
 The same happens if we choose $V_{\tilde{N}-k}$ instead. In that
case, the equation above is $\mathbf{M}V_{\tilde{N}-k}=V_{\tilde{N}+k}$
as expected. A matrix which has exactly the same action as above can
be found easily and this is $\left(\begin{array}{cc}
 0 & \lambda _{1}/2\\
 \lambda _{1}/2 & 0\end{array}
\right)$. Finally, we observe that

\[
\left(\begin{array}{cc}
 \left(\frac{\tilde{N}+k}{L}\right)^{2}+\frac{\lambda _{1}}{2}-\lambda _{2}-\omega ^{2} & \frac{\lambda _{1}}{2}\\
 \frac{\lambda _{1}}{2} & \left(\frac{\tilde{N}-k}{L}\right)^{2}+\frac{\lambda _{1}}{2}-\lambda _{2}-\omega ^{2}\end{array}
\right)V_{\tilde{N}+k}=0\]
 which implies the condition that the determinant of the above matrix
must be zero in order to acquire the non-zero eigenvalues.

The eigenvalues of $A\left(\tilde{N},L\right)$ on $V_{n}$ are

\begin{align}
\omega ^{2}\left(N,k,\pm 1\right)= & \frac{1}{L^{2}}\: \: \left[L^{2}-\left(N+\frac{1}{2}\right)^{2}+k^{2}\right]\nonumber \\
\pm  & \frac{1}{L^{2}}\, \sqrt{\left(L^{2}-\left(N+\frac{1}{2}\right)^{2}\right)^{2}+4k^{2}\left(N+\frac{1}{2}\right)^{2}}
\end{align}
for $k=0,1,2,...$ and similarly on $\tilde{V_{n}}$ for $k=1,2,3,...$
. Notice that if we replace $N+1/2$ by $\tilde{N}$ then the above
result is the same with eq.(36) of {[}1{]}.

For $N$ having a specific value and $L$ slightly greater than $N$,
the solution (10)-(11) possess $4N$ negative modes corresponding
to $k=1,2,3,...,2N$ in (37) (minus sign). When $L$ increases, the
number of positive eigenmodes increases as well

\begin{equation}
L^{2}=\frac{3}{4}\left(2N+1\right)^{2}-\frac{m^{2}}{2}\! ,\quad \quad \quad m=1,2,...,2N,\end{equation}
 i.e. (cf(23)) at those values of $L$ where the solutions of type-III
with $n=N$ bifurcate from the solution of type-I. For 

\begin{equation}
L^{2}\geq L_{cr}^{2}\left(N\right)\equiv \frac{3}{4}\left(2N+1\right)^{2}-\frac{1}{2}\end{equation}
 all the modes are positive and (10)-(11) are classically stable solitons.
These results are illustrated in figure 2 for $N=1$ and $N=2$. The
numbers in parentheses represent the number of negative and zero modes
of the corresponding branch.

\subsection{{\normalsize Type-II and Type-III solutions}}

The equations for the fluctuations about the solution (13) decouple
to take the form of Lame equations:

\begin{equation}
\left\{ -\frac{d^{2}}{dy^{2}}+6k^{2}sn^{2}\left(y,k\right)\right\} \eta _{1}=\Omega _{1}^{2}\eta _{1}\end{equation}

\begin{equation}
\left\{ -\frac{d^{2}}{dy^{2}}+2k^{2}sn^{2}\left(y,k\right)\right\} \eta _{2}=\Omega _{2}^{2}\eta _{2}\end{equation}
 where

\begin{equation}
x=\sqrt{1+k^{2}}\: y\: ,\quad \quad \Omega _{a}^{2}\equiv \left(\omega _{a}^{2}+1\right)\left(k^{2}+1\right)\! ,\quad \quad a=1,2\end{equation}
 and $\omega _{a}^{2}$ is the effective eigenvalue of the relevant
operator. Equations (40) and (41) admit two and one algebraic modes,
respectively, with corresponding eigenvalues

\begin{equation}
\Omega _{1}^{2}:\: 4+k^{2}\: ,\: 1+4k^{2}\qquad \quad and\qquad \quad \Omega _{2}^{2}:\: 1+k^{2}\end{equation}
The corresponding values of $\omega ^{2}$ follow immediately from
(42); they have signature (+,+) and (0), respectively.

It's a property of the Lame equation that the solutions determined
algebraically correspond to the solutions of the lowest eigenvalues.
The remaining part of the spectrum therefore consists of positive
eigenvalues. The spectrum of the equation (40) was studied perturbatively
in {[}9{]}, while the relation between the Lame equation and the Manton-Samols
sphalerons was first pointed out in {[}11-13{]}. 

Concerning the stability equation of type-III solution, it seems to
be more difficult to deal with as the presence of the Jacobi Elliptic
functions prevent us from having an analytic expression for $\Omega (x)$
in (4). Thus, we are not able to follow the steps done in the case
of type-I solutions (i.e. find the \textbf{M} matrix) where trigonometric
functions are easier to handle. Detailed analysis especially for the
case of Lame equations can be found in {[}12{]} and {[}15{]}.

\section{{\normalsize Conclusion}}

A detailed analytical study of the static solutions of the Goldstone
model on a circle has been given in {[}1{]} and we followed the same
path here for our boundary conditions. Many results of {[}1{]} are
connected with ours by a simple change on variables used. We write
them explicitly above whenever is necessary. We also note our effort
to impose mixed boundary conditions as well. Specifically, we enforced
$\phi _{1}$ to be antiperiodic and $\phi _{2}$ periodic but there
was no solution to satisfy this choice so it's needless to extend
beyond this small remark here.

Many details were presented on the stability analysis on section 3
for the solutions found. Classically stable solitons were identified,
together with the range of the parameter $L$ for which they are stable. 

This simpler model we present here and many others can be connected
and can also give us the experience to deal with realistic (3+1)-dimensional
particle physics models in our search for possible metastable localized
solitons. Ref. {[}1{]} on which this note was based, has connections
with {[}9{]} and {[}10{]} as it concerns the branches of their solutions,
also with {[}2-6{]}, while there are also interesting physical applications
{[}14{]} as it is already mentioned in {[}1{]}. Soliton solutions
in (3+1)-dimensional models with antiperiodic boundary condition imposed
on one spatial dimension, which is compactified on $S^{1}$, is analyzed
in {[}16-18{]} where supersymmetry breaking is found.

\section*{{\normalsize Acknowledgements}}

The author is very thankful to Professor T. N. Tomaras for fruitful
conversations and useful advice. Work supported in part by the EU
contract HPRN-CT-2000-00122.

\end{document}